**Title:** Low frequency fringe pattern analysis *via* a Fourier transform method.

**Authors:** Andrew John Henning, email: a.henning@hud.ac.uk

Dawei Tang, email: d.tang@hud.ac.uk

Xiangqian (Jane) Jiang, email: x.jiang@hud.ac.uk

**Affiliation:** EPSRC future metrology hub, University of Huddersfield, Queensgate, Huddersfield, HD1 3DH

**Abstract:** The analysis of signals created by a variety of instruments involves calculating the phase of a sinusoidal type signal. One widely used method to extract this information is through the use of Fourier transforms, but it is known that significant errors can arise when a low number of cycles of a sinusoid are present in the signal. In the following, we examine the case where the fringe pattern of interest only contains a few cycles of a sinusoid, looking at the adjustments that need to be made to the previous method in order to allow it to be used successfully, the causes of errors that arise, and the accuracy that can be expected to be obtained.

**Introduction**

It is not sufficient just to have an instrument that generates a stable signal in a well defined manner, but this needs to be combined with an analysis of the signal that yeilds an accurate value of the desired parameter. A very successful and a widely used technique utilising Fourier transforms (FT) was introduced by Takeda *et al*[1] to accurately recover the phase from sinusiodal type signals. This method has been applied to a broad range of measurement scenarios including: off axis holography[2], X-ray Talbot interferometry[3], Stimulated Raman scattering[4], Quantitative phase imaging[5], Fourier domain optical coherence tomography[6], Wavelength scanning interferometry[7,8], high resoloution electron microscopy[9], Shearography[10], velocity measurments[11] and many others. While providing highly accurate results for high frequency signals, the accuracy of the method degrades for low frequency signals.

In their original paper[1] the authors looked at signals consisting of a high frequency carrier signal modulated by a slowly varying envelope, and with a slowly varying background signal. However, in many applications lower frequency signals are encountered, and so a lower bound to the applicability of the method needs to be found. In addition it can be shown that, with slight modifications to the application of the method, a good result for the phase of the signal can be obtained even when the signal only forms a few cycles of a sinusoid across the detector.

In the following we establish the causes of the significant errors that can arise when the method is applied to low frequency signals. We demonstrate that the errors are fundamental to the method, and do not arise from noise or other signal degradation, and even occur when there is no envelope modulating the signal. Using this analysis we present simple adaptations to the previous method, such as initially subtracting the background signal, and modifying the envelope on the interference fringes to reduce the width of its FT, that should be applied when low frequency signals are considered. We look at the size of errors that can be expected for different ranges of low frequency signals, show how the expected error can be calculated from the envelope on the signal and how the FT of the envelope can be used to suggest a lower limit to the frequency that this method should be used for in order to provide a given accuracy. We subsequently verify the assertions experimentally, demonstrating that the phase is well recovered for several low frequency signals.

**Results**

The method presented in ref 1. looks at signals that take the quite general form of

$$s_t\{x,y\} = a\{x,y\} + b\{x,y\}\cos(k_0 x + \phi\{x,y\}) \qquad (1)$$

where $s_t\{x,y\}$ is the signal recorded, $a\{x,y\}$ is the background signal, $b\{x,y\}$ is the envelope of the interference fringes, $k_0 = 2\pi/\lambda$ is the wavenumber, $\phi\{x,y\}$ is an additional phase term, and $\lambda$ is the wavelength of the signal recorded on an $x - y$ plane. For simplicity we will look at the signal in one dimension, which will be a line along the $x$ axis, as this is the result that is required for the instruments that we use, and the generalization to 2 dimensions is trivial and all of the ideas that follow still hold. Thus eq. 1 becomes

$$s\{x\} = a\{x\} + b\{x\}\cos(k_0 x + \phi\{x\}) \qquad (2)$$

This type of signal is illustrated in fig. 1(a). This signal is typical for a variety of interferometric instruments where there is a constant envelope on the fringes that can be attributed to such things as variations in the intensity of the source, and a background signal due to light that does not interfere, due to e.g. different intensity of light from the measurement arm and reference arm, or background light falling on the detector.

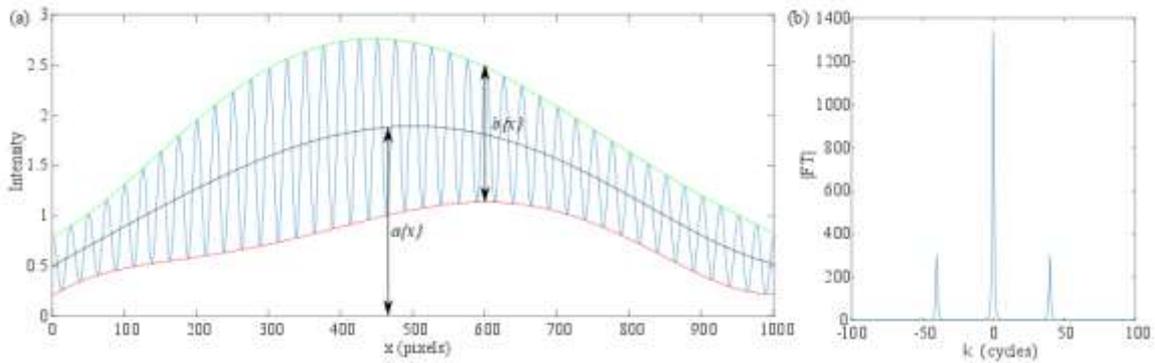

Figure 1: Part (a) shows a typical signal of the type generated from the interference of light. $a\{x\}$ corresponds to a background signal, such as stray light falling onto the detector, or incoherent light passing through the system, while the interference fringes are modulated by a function $b\{x\}$, which may be due to such things as variation of the sources intensity across the spectrum. Part (b) shows the modulus of the signals FT.

Although $k_0$ is a constant, as $\phi\{x\}$ is arbitrary the cosine term in eq. 2 has no fixed period which complicates matters. However, as the function is defined over a limited range along the $x$ axis, corresponding to the length of the detector it is recorded on or time period over which the signal is recorded, we can make the substitution

$$\cos(k_0 x + \phi\{x\}) = \sum \alpha_n \cos(k_n x + \theta_n) \qquad (3)$$

where the r.h.s. of eq. 3 is just the Fourier series representation of the signal, each term of which has a fixed frequency and phase term. As the following steps are all linear, we can deal with each term separately and sum the individual results. For simplicity, we will start by looking at a signal where

$\cos(k_0 x + \phi\{x\}) = \cos(k_s x + \theta_s)$, where $k_s x$ is the wavenumber of a single term in eq. 3, thus all of the other terms will be equal to zero.

If we convert the cosine term into its exponential form, as in ref 1.,

$$s\{x\} = a\{x\} + \frac{b\{x\}}{2}\exp(i\theta_s)\exp(ik_s x) + \frac{b\{x\}}{2}\exp(-i\theta_s)\exp(-ik_s x) \tag{4}$$

assuming that $a\{x\}$ and $b\{x\}$ are slowly varying functions, it can be seen that the FT of the signal contains three peaks, one around the $k = 0$ axis, and one around each of the points $k = \pm k_s$, as shown in fig 1(b). The FT of eq. 4 can be viewed as the sum of the FT of each of the three terms, and we assume that the signal around $k_s$ in the FT is solely due to the second term in eq. 4.

We will initially further simplify the problem by setting $a\{x\} = 0$ and $b\{x\} = 1$ which corresponds to removing the background component of the signal and the envelope around the interference fringes, before subsequently looking at more general forms of $a\{x\}$ and $b\{x\}$ to see how they affect the phase determination. Applying these simplifications leaves a sinusoidal signal scaled between $\pm 1$.

$$s_1\{x\} = \frac{1}{2}\exp(i\theta_s)\exp(ik_s x) + \frac{1}{2}\exp(-i\theta_s)\exp(-ik_s x) \tag{5}$$

With the FT of this signal, $S_1\{k\}$, being

$$S_1\{k\} = \frac{C}{2}\exp(i\theta_s)\delta(k_s - k) + \frac{C}{2}\exp(-i\theta_s)\delta(k_s + k) \tag{6}$$

where $\delta$ is a Dirac delta function, and $C$ is a scaling constant. Taking the inverse FT (IFT) of the first term on the r.h.s. of eq. 6 gives

$$\text{IFT\{term 2\}} = \frac{1}{2}\exp(i(k_s x + \theta_s)) = \frac{1}{2}[\cos(k_s x + \theta_s) + i\sin(k_s x + \theta_s)] \tag{7}$$

The real part of the exponential term in eq. 7 has the same magnitude at all values of $x$ as the $\cos(k_s x + \theta_s)$ term of the initial signal, but an imaginary term has been introduced that has the magnitude of the sine of the signal, thus the phase is given by the angle on the complex plane between the vector $\exp(i(k_s x + \theta_s))$ and the real axis. If the peak that is on the negative frequency axis is retained instead of the peak on the positive axis, i.e. the second term in eq. 6 instead of the first, the result of the inverse Fourier transform (IFT) gives a vector with same real component but changes the sign of the imaginary component.

The above immediately answers the question of what the recovered phase means in the case where the frequency of the cosine term is not constant as, following eq. 3, we can see the answer is the angle that the vector sum of the result for each of the individual frequency terms makes with the real axis on the complex plane. It should be noted that in the paper by Takeda et al[1], the process they use is to take the complex log of the IFT, and then the phase is given as the imaginary component of the result, however this is equivalent to taking the angle of the complex number in polar form.

<u>The origin of the low frequency errors</u>

The signal used in the previous section does not quite correspond to the case for a true experimental result which has only a limited extent, and is recorded at discrete points. In this section we will show that the limited length of the signal means that errors will occur at low frequencies even for a perfect signal when using the method described in the previous section.

In the following we will consider the case where this signal is recorded on a 1001 pixel camera, which extends from $x = -L/2$ to $x = L/2$. This effect can be introduced *via* the $b\{x\}$ term, by setting

$$b\{x\} = \begin{cases} 0, & x < L/2 \\ 1, & -L/2 \leq x < L/2 \\ 0, & x \geq L/2 \end{cases} \tag{8}$$

The Fourier transform of the recorded signal is thus the Fourier transform of $s_1\{x\}b\{x\}$, which by the convolution theorem is $\text{FT}[s_1\{x\}b\{x\}] = \text{FT}[s_1\{x\}] \otimes \text{FT}[b\{x\}]$.

$$\begin{aligned}\text{FT}[s_1\{x\}b\{x\}] &= \left(\frac{C \exp(i\theta_s)}{2}\delta(k_s - k) + \frac{C \exp(-i\theta_s)}{2}\delta(k_s + k)\right) \otimes L\text{sinc}\left[\frac{kL}{2}\right] \\ &= \frac{CL \exp(i\theta_s)}{2}\text{sinc}\left[\frac{(k - k_s)L}{2}\right] + \frac{CL \exp(-i\theta_s)}{2}\text{sinc}\left[\frac{(k + k_s)L}{2}\right]\end{aligned} \tag{9}$$

When there is a whole number of cycles across the detector, $k_s$ corresponds to one of the frequencies in the discrete Fourier transform (DFT), and the sinc term in the FT of the window is equal to zero at all of the other recorded frequencies. However, when $k_s$ lies between the frequencies of the terms in the DFT, the sinc function leads to the data from the $\exp(\pm ik_s x)$ terms being spread along the $k$ axis when it is convolved with the FT of the window function. As the sinc function does not decay particularly fast, this can introduce terms of significant magnitude at numerous points along the axis. As we are considering this in terms of number of cycles across the detector, the results are general to any length of detector.

When the method was explained in the previous section, it was noted that the signal is FT'd before either the positive or negative set of frequencies are set to zero, before an inverse FT is carried out. The spreading of the signal along the $k$ axis by the sinc function means, however, that the signal that would be confined to the positive section of the spatial frequency axis for an infinite signal is spread into the negative frequency axis, and vice-versa. Thus when one half of the axis is set to zero, some of the information that we wish to retain is lost, while some of that which should be removed is retained.

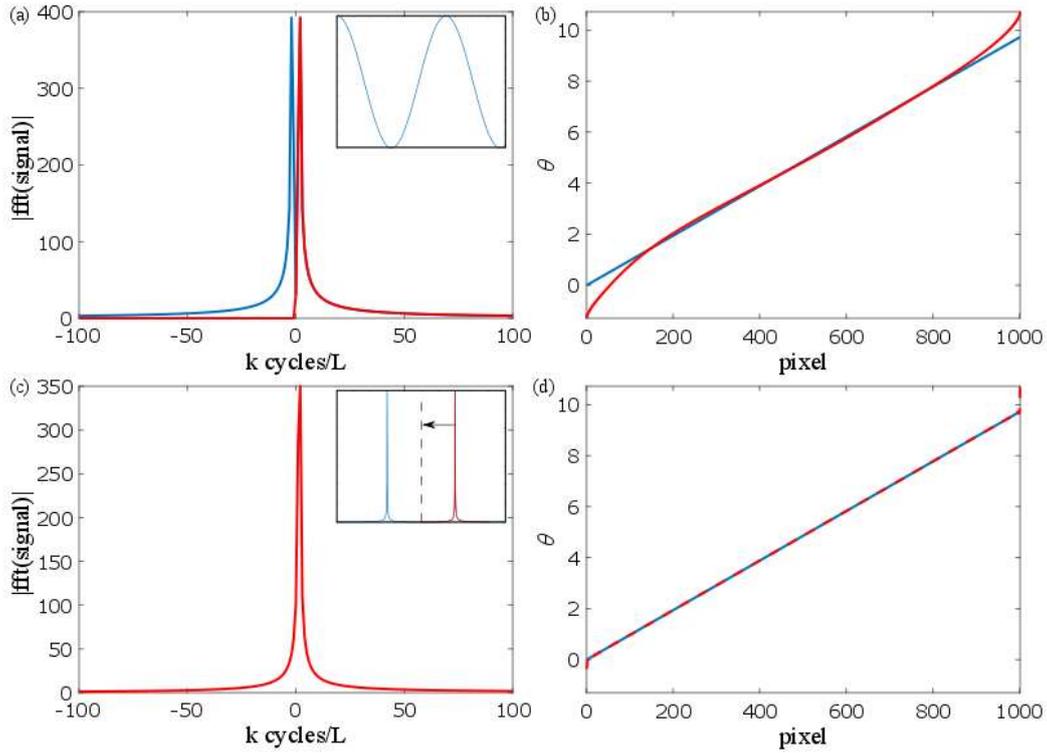

Figure 2: Part (a) shows the magnitude of the FT of a signal of 1.55 cycles across the detector (shown in inset). The section outlined in red is the data retained when the negative frequencies are set to zero. Part (b) shows the true phase (blue line), and the recovered phase using the positive frequency data. Part (c) shows the magnitude of the FT generated by shifting a single peak from a signal with 201.55 cycles across the detector shifted along the axis as shown in the inset. Part (d) shows the phase recovered from this spatial frequency signal shown in part (c), (red dashed line), and the true phase (blue line).

This fact is illustrated in fig. 2. A signal corresponding to $\cos(\theta)$ where $0 < \theta < 1.55(2\pi)$, is generated, as shown in the inset of fig. 2(a). The main section of part (a) shows the FT of this signal, with the red line marking the data that is retained when the data on the negative frequency portion of the axis is set to zero. In part (b) the blue line corresponds to the value of $\theta$ at each point used to generate the signal, while the red line shows the phase recovered when taking the phase angle of the IFT of the retained spatial frequency data.

While we could calculate the result of convolving the FT of the window with the single $\delta$ function directly, we can achieve a similar result from the FT of a higher frequency signal, which makes it clear that the difference in the result is not due to the different ways in which the FT's were generated. To achieve this a signal with an integer number of additional cycles is generated, in this case $201.55$ cycles instead of 1.55. The FT of this produces two well separated peaks, which can once again be looked at as the convolution of the FT of the window function, $B\{k\}$, with $S_1\{k\}$ from eq. 6. As the fractional part of the number of cycles is the same for the two cases, the $\delta$ function lies at the same point between the 201 and 202 cycle points in the DFT as the $\delta$ function for the lower frequency signal does between the 1 and 2 cycle points. Thus, ignoring the second peak, the result at the recorded points in the DFT for the two cases is the same, just translated along the axis. By retaining the data around one peak and translating it along the spatial frequency axis to remove the additional number of cycles, the signal that is due to the broadening of a single $\delta$ function can be obtained. This method we have implemented here to create the lower frequency signal from the higher frequency signal is the same as is used in ref 1 where the retained signal is shifted to the origin in order to remove the carrier signal.

The FT of this higher frequency signal is shown in the inset of fig 2(c). The data marked in red is that which is retained, all of the rest of the data is set to zero, and the data marked in red is shifted 200 cycles along the spatial frequency axis towards the origin. The phase recovery method is applied to this signal, with the results being shown by the red dashed line in part (d), which closely matches the blue line corresponding to the phase used to generate the initial sinusoidal signal.

This shows that we have discarded information that we would wish to retain when the peaks are close to the $k = 0$ axis, however, there is also information from the broadened $\delta$ fn from the negative frequency axis that is retained on the positive set of frequencies. The way in which the phase recovered is changed can be seen more clearly by looking at the elements in the FT that are retained in the case shown in part (a) of fig 2, and those retained in part (c).

The retained terms for the case shown in fig. 2(a) are

$$S_a\{k\} = \frac{CL}{2}\left(\exp(i\theta_s)\mathrm{sinc}\left[\frac{(k-k_s)L}{2}\right] + \exp(-i\theta_s)\mathrm{sinc}\left[\frac{(k+k_s)L}{2}\right]\right), \qquad k \geq 0$$

$$= \frac{CL}{2}\left((\alpha + i\beta)\mathrm{sinc}\left[\frac{(k-k_s)L}{2}\right] + (\alpha - i\beta)\mathrm{sinc}\left[\frac{(k+k_s)L}{2}\right]\right), \qquad k \geq 0$$

$$= \frac{CL}{2}\left(\alpha\left(\mathrm{sinc}\left[\frac{(k-k_s)L}{2}\right] + \mathrm{sinc}\left[\frac{(k+k_s)L}{2}\right]\right)\right.$$
$$\left. + i\beta\left(\mathrm{sinc}\left[\frac{(k-k_s)L}{2}\right] - \mathrm{sinc}\left[\frac{(k+k_s)L}{2}\right]\right)\right) \qquad k \geq 0 \quad (10)$$

Whereas in the case shown in fig 2(c) the retained terms are

$$S_c\{k\} = \frac{CL}{2}(\alpha + i\beta)\mathrm{sinc}\left[\frac{(k-k_s)L}{2}\right] \qquad (11)$$

Looking at the real parts of eqn's 10 and 11, it can be seen that, while the negative frequency elements in eq. 10 have been reduced to zero, all of the positive frequency elements magnitudes have an additional element due to the $\mathrm{sinc}\left[\frac{(k+k_s)L}{2}\right]$ term, which for each value of $k$ is equal to the magnitude of the element lost at $-k$. However, when looking at the imaginary parts it can be seen that the additional element is of equal magnitude but opposite sign to the element lost.

To examine the effect that the difference between $S_a\{k\}$ and $S_c\{k\}$ has on the phase recovered, we need to look at their IFT's. In the case where both peaks are present the result is

$$\mathrm{IFT}[S_a\{k\}] = \frac{CL}{2}\int_0^\infty \mathrm{sinc}\left[\frac{(k+k_s)L}{2}\right]\exp(-i\theta)\exp(ikx) + \mathrm{sinc}\left[\frac{(k-k_s)L}{2}\right]\exp(i\theta)\exp(ikx)\,dk$$

$$= \frac{CL}{2}\int_0^\infty \mathrm{sinc}\left[\frac{(k+k_s)L}{2}\right](\alpha - i\beta)\exp(ikx) + \mathrm{sinc}\left[\frac{(k-k_s)L}{2}\right](\alpha + i\beta)\exp(ikx)\,dk \quad (12)$$

While for the single peak case

$$\text{IFT}[S_c\{k\}] = \frac{CL}{2}\int_{-\infty}^{\infty} \text{sinc}\left[\frac{(k-k_s)L}{2}\right]\exp(i\theta)\exp(ikx)dk$$

$$= \frac{CL}{2}\int_{-\infty}^{0} \text{sinc}\left[\frac{(k-k_s)L}{2}\right]\exp(i\theta)\exp(ikx)dk + \int_{0}^{\infty} \text{sinc}\left[\frac{(k-k_s)L}{2}\right]\exp(i\theta)\exp(ikx)dk$$

$$= \frac{CL}{2}\int_{0}^{\infty} \text{sinc}\left[\frac{(k+k_s)L}{2}\right](\alpha+i\beta)\exp(-ikx) + \text{sinc}\left[\frac{(k-k_s)L}{2}\right](\alpha+i\beta)\exp(ikx)\,dk \quad (13)$$

Writing $\exp(ikx)$ as $e_r\{k\} + ie_i\{k\}$, and separating the real and imaginary components gives

$$\text{IFT}[S_a\{k\}] = \frac{CL}{2}\int_{0}^{\infty}\left[(\alpha e_r + \beta e_i)\text{sinc}\left[\frac{(k+k_s)L}{2}\right] + (\alpha e_r - \beta e_i)\text{sinc}\left[\frac{(k-k_s)L}{2}\right]\right.$$
$$\left. + i\left[(\alpha e_i - \beta e_r)\text{sinc}\left[\frac{(k+k_s)L}{2}\right] + (\alpha e_i + \beta e_r)\text{sinc}\left[\frac{(k-k_s)L}{2}\right]\right]\right]dk \quad (14)$$

And

$$\text{IFT}[S_c\{k\}] = \frac{CL}{2}\int_{0}^{\infty}\left[(\alpha e_r + \beta e_i)\text{sinc}\left[\frac{(k+k_s)L}{2}\right] + (\alpha e_r - \beta e_i)\text{sinc}\left[\frac{(k-k_s)L}{2}\right]\right.$$
$$\left. + i\left[(\beta e_r - \alpha e_i)\text{sinc}\left[\frac{(k+k_s)L}{2}\right] + (\alpha e_i + \beta e_r)\text{sinc}\left[\frac{(k-k_s)L}{2}\right]\right]\right]dk \quad (15)$$

It can be seen that $\text{IFT}[S_a\{k\}]$ is the same as $\text{IFT}[S_c\{k\}]$ except for the coefficient of the $\text{sinc}\left[\frac{(k+k_s)L}{2}\right]$ term in the imaginary part of the integral.

We can convert eqns. 14 and 15 into the result that would be used obtained by a discrete IFT (DIFT) of the signal result by multiplying the terms in the integral by a comb of $\delta$ functions, and then replacing the integral by the sum. The delta functions are centred at points on the spatial frequency axis that correspond to a whole number of cycles over the detector length. Equation 14 then becomes

$$\text{DIFT}[S_a\{k\}] = \sum_{m=0}^{n}\left[(\alpha e_r + \beta e_i)\text{sinc}\left[\frac{(k\{m\}+k_s)L}{2}\right] + (\alpha e_r - \beta e_i)\text{sinc}\left[\frac{(k\{m\}-k_s)L}{2}\right]\right.$$
$$\left. + i\left[(\alpha e_i - \beta e_r)\text{sinc}\left[\frac{(k\{m\}+k_s)L}{2}\right] + (\alpha e_i + \beta e_r)\text{sinc}\left[\frac{(k\{m\}-k_s)L}{2}\right]\right]\right] \quad (16)$$

Where $k\{m\} = 2\pi m/L$. The phase is then given by the angle the vector sum makes with the positive real axis. Changing the imaginary part of each of the elements without changing the real part is likely to change the angle of the vector sum, and will be the cause of the phase errors.

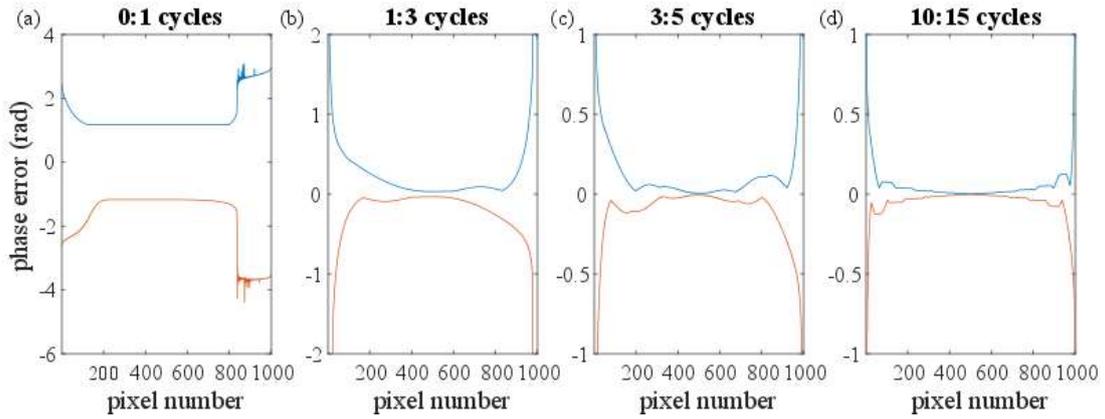

Figure 3: The upper and lower bounds to the errors found over several frequency ranges. The blue and red lines mark the greatest difference between the phase recovered and the phase used to generate a signal with (a) 0-1, (b)1-3 (c)3-5 and (d) 10-15 cycles across the detector. It can be seen that as the number of cycles increases so does the accuracy of the method, and that the most accurate results are found in the centre of the measurement array.

The typical size of the phase errors that can be expected over different ranges are illustrated in fig. 3. The errors are calculated by generating a set of sinusoidal signals, and then applying the method outlined earlier to recover the phase from the signal, with the error being the difference between the two phases. The results are calculated for every 0.1 cycle within the stated range, and the phase offset of the signal is varied. In part (a) the upper and lower limit in the range 0-1 cycles is found, part (b) 1-3 cycles, part (c) 3-5 cycles and part (d) 10-15 cycles. It can be seen that the phase errors are worse for fewer numbers of cycles, especially at the edge of the range, though once a few cycles are present the errors become reasonably small. The importance of these errors will depend upon the accuracy required, but it is clear that the recovered phase values from the centre of the signal are likely to give a more accurate result.

Effects of an envelope on the signal

In the previous section, the envelope on the signal was limited to $b\{x\} = 1$ in the range $-L/2 \leq x \leq L/2$, this however is highly unlikely to be true in a real instrument. Most envelopes are of the type illustrated in fig. 1, where the magnitude of the fringes changes smoothly and slowly. This implies that its FT will be dominated by the low frequency terms, and should drop off fairly rapidly. Rapid changes in the envelope imply a significant magnitude for some of the higher frequency terms of $B\{k\}$, which will lead to significant phase recovery errors.

As shown by eq. 9, the FT of the signal modulated by the envelope can be viewed as the convolution of the FT of the envelope, $B\{k\}$, with the FT of the infinite sinusoidal signal. The convolution of $B\{k\}$ with the two $\delta$ functions produces two shifted copies of $B\{k\}$, one around $k_s$ and the other around $-k_s$. These shifted functions are multiplied by the other coefficients of the respective delta function, but the phase errors arise in a similar manner as in the case where $b\{x\} = 1$ within the measurement range, though as the form of $B\{k\}$ is changed, so are the errors that arise.

To look at the effect of a general window on the signal we start with the single broadened peak around $k = k_s$. The convolution of the complex valued FT of the window, $B\{k\}$, with the term 2 in eq. 6 gives

$$S_d\{k\} = \frac{C}{2}(\alpha + i\beta)B\{k - k_s\} = \frac{C}{2}(\alpha + i\beta)(\gamma\{k - k_s\} + i\epsilon\{k - k_s\})$$
$$= \frac{C}{2}[(\alpha\gamma\{k - k_s\} - \beta\epsilon\{k - k_s\}) + i(\alpha\epsilon\{k - k_s\} + \beta\gamma\{k - k_s\})] \quad (17)$$

This can be combined with the fact that we know that $b\{x\}$ is a real valued function meaning that $B\{k\} = B^*\{-k\}$, thus for each frequency $k = k_s + k_a$ where $B\{k_s + k_a\} = \gamma\{k_a\} + i\epsilon\{k_a\}$ there is a frequency $k = k_s - k_a$ at which $B\{k_s - k_a\} = \gamma\{k_a\} - i\epsilon\{k_a\}$. The sum of these two terms is

$$S_d\{k_s + k_a\} + S_d\{k_s - k_a\} = \gamma\{k_a\}C[\alpha + i\beta] \quad (18)$$

Thus the phase of the sum of these two terms is the same as the phase of the $\exp(i\theta)$ term, however the magnitude of the sum of these pair of terms is changed by an amount $\gamma\{k_a\}$. This is true for all values of $k_a$, and as the sum of each pair of vectors forms a vector in the direction $\alpha + i\beta$, the angle of the vector sum is unchanged. The $k = 0$ term in $B\{k\}$ is real, and so, while it does not have a pair, when convolved with the $\delta$ function still gives a vector in the direction $\alpha + i\beta$.

As with the previous section, this is not the signal that we retain when the FT is taken and the negative frequency components are removed as, like before, this removes some of the elements related to the convolution with the $\delta$ on the positive frequency axis, meaning that some $S_d\{k_s + k_a\}$ terms have lost their pairs at $S_d\{k_s - k_a\}$, and there are additional terms retained from the convolution of the window with the $\delta$ fn terms on the negative frequency axis.

Again, due to the fact that $B\{k\} = B^*\{-k\}$, it can be seen that the elements lost are

$$S\{-k\}_{lost} = \frac{C}{2}(\alpha + i\beta)B\{-k\} = \frac{C}{2}[(\alpha\gamma\{k\} - \beta\epsilon\{k\}) + i(\alpha\epsilon\{k\} + \beta\gamma\{k\})] \quad k > 0 \quad (19)$$

While those gained are

$$S\{k\}_{gained} = \frac{C}{2}(\alpha - i\beta)B\{k\} = \frac{C}{2}[(\alpha\gamma\{k\} - \beta\epsilon\{k\}) - i(\alpha\epsilon\{k\} + \beta\gamma\{k\})] \quad k > 0 \quad (20)$$

Where, as before, for each element of the FT lost at a value of $-k$, there is an additional element added to the $k$ term whose real part is the same as that lost and whose imaginary part is of the same magnitude but opposite sign. The only difference with the previous case is that the FT of the window has a non-zero imaginary part which affects the phase angle of the unpaired components.

Should the envelope be smooth and slowly varying, it is likely that the low frequency terms dominate, and thus the magnitude of the terms lost and gained are likely to be small, leading to a small change in the phase angle recovered. The wider the region over which terms in the FT of the window have significant magnitude, the higher the limit to the frequency at which the phase recovery will give useable results for any given accuracy.

<u>Effects of the background signal</u>

When the phase recovery method is described in Takeda *et al*[1], it is noted that the high frequency of the signal which we are interested means the FT of the signal will form 3 separate peaks, one around $k = \pm k_s$ and one around $k = 0$. By eliminating the peak around $k = 0$ the background signal, that due to $a\{x\}$ in eq. 4, is removed. However, as we push the frequency of the signal to lower frequencies, this is no longer possible.

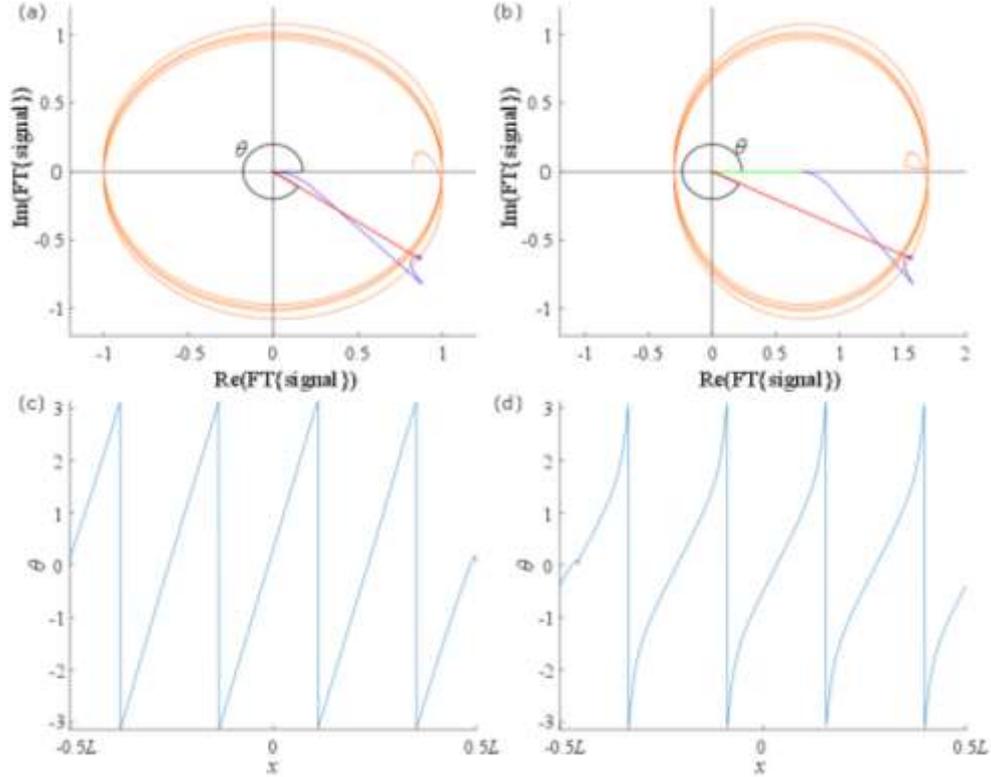

Figure 4: Part (a) shows the real and imaginary parts of the FT of a sinusoidal signal with 4.1 cycles across the detector. The dark blue line shows the vectors corresponding to each of the elements of the positive frequencies of the FT for a single x position, while the red line is a vector corresponding to the sum of all of the elements. The orange line traces the end of the vector throughout the x range. Part (b) shows the same but where a DC term, shown by the green vector has been added. This shifts the set of solutions and changes the angle recovered. Part (c) shows the phase recovered for the case in part (a), while part (d) shows the phase for the case in part (b).

Returning to the full expression for the signal, given by eq. 4, and taking the FT gives

$$S\{k\} = \int_{-\infty}^{\infty} a\{x\} \exp(-ikx)\, dx + \int_{-\infty}^{\infty} \frac{b\{x\}}{2} \exp(i\theta_s) \exp(ik_s x) \exp(-ikx)\, dx$$
$$+ \int_{-\infty}^{\infty} \frac{b\{x\}}{2} \exp(-i\theta_s) \exp(-ik_s x) \exp(-ikx)\, dx \quad (21)$$

the second and third term on the r.h.s. can be dealt with just as has been done so previously, but the term including $a\{x\}$ will be added onto the result. The effect of this on the phase recovered is illustrated in fig. 4. In part (a) the blue curve shows each of the vectors in the discrete IFT, $\sum S\{k\} \exp(-ikx)$, for $S\{k\}$ where $k \geq 0$, for a particular value of $x$. The red line is just the vector from the origin to the end of the vector sum, and it is the angle that this vector makes with the real axis that is the phase returned by the phase recovery method. The orange line marks the locus of points that the end of the red vector traces out as $x$ is varied from $-L/2$ to $L/2$. In part (a), $a\{x\} = 0$, while in part (b) the effect of $a\{x\}$ being equal to a non-zero constant is shown. The presence of $a\{x\}$ then increases the $k = 0$ term in $S\{k\}$, shifting the end point of the vector sum along the real axis and modifying the phase recovered. Parts (c) and (d) show the phase recovered across the range $-L/2 \leq x \leq L/2$, for the cases in parts (a) and (b) respectively. The translation of the locus of end points can be seen to significantly modify the phase recovered. If the $a\{x\}$ term takes a more complex form than a constant background signal then more terms will be modified, introducing further errors. One way of looking at this is that the recovered phase becomes a mixture of the phase of the signal that we are interested in, and the phase of the background signal.

It should be noted that when it is an intensity recorded on the detector, all signals will need to be shifted to put the upper and lower bounds of the envelope equal about the $y$ axis before the phase recovery method is applied, as $A\{k\}$ will always contain a non-zero DC term.

The errors from the background terms, $a\{x\}$, are likely to be highly significant, and so if the phase recovery method is to be used for low frequency signals then it is likely that the background signal must be well characterised and removed before the signal is FT'd. The middle of the envelope on the signal, $b\{x\}/2$, should be shifted to lie on the $y = 0$ axis. Should the envelopes vary throughout a measurement in a way that cannot be characterised it is likely that this algorithm is likely to yield poor results.

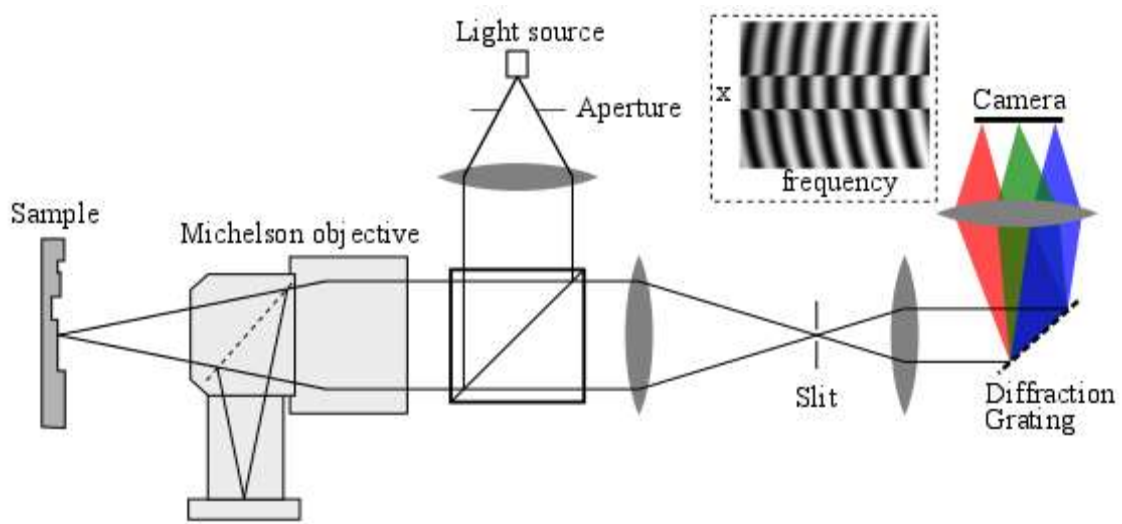

Figure 5: A schematic of the instrument that the experimental results are taken from. The instrument measures a line profile on the sample, and for each point along the profile the white light from the light source is split into its separate wavelength components, forming a signal of the type illustrated in the inset. The signal generated for each point along the profile forms a sinusoid whose frequency is related to the distance of the sample from the point of equal optical path length between the reference path and the measurement path.

Experimental verification

In this section we look at the ideas in the previous section applied to data that is generated by the instrument illustrated in fig. 5. Light from an extended, spatially incoherent, source is focused and split by a 4x Michelson objective with a working distance of 27mm, with light in one path scattering from a reference mirror, while light in the other path scattering from the object being measured. The light from each path is collected and imaged onto a slit only allowing light from a line to pass, meaning that the signal relates to light from a line on the object and reference mirror. This is collimated before being incident on a diffraction grating, which separates the light by wavelength in the direction perpendicular to the measurement profile. This is focussed by a lens onto a camera (IMPERX, Bobcat B06204,480 by 640 pixels) producing a signal in which each row of pixels corresponds to the interference between light from a different pair of points on the reference surface and object being measured, and each column corresponds to the interference of the light at different wavelengths. Ideally the wavenumber varies linearly with pixel number the signal forms a sinusoid along the frequency axis whose period is related to the optical path difference between the measurement and reference arm at the points being measured. By measuring the frequency of the sinusoid the location of each point along the profile can be found. A typical result is illustrated in the inset in fig. 5, where the sinusoidal signal along the frequency axis different for the rows corresponding to the $x$ positions in the centre of the range. This is because there is a groove on the object being measured. In the true experimental case the wavenumber does not change perfectly linearly along the frequency axis. For this reason, the instrument was calibrated and the data resampled from the 640 measured points onto set of 1001 points the are equally spaced w.r.t. wavenumber, between the maximum and minimum wavenumber recorded. This then matches number of measurement points used in the numerical examples in previous sections.

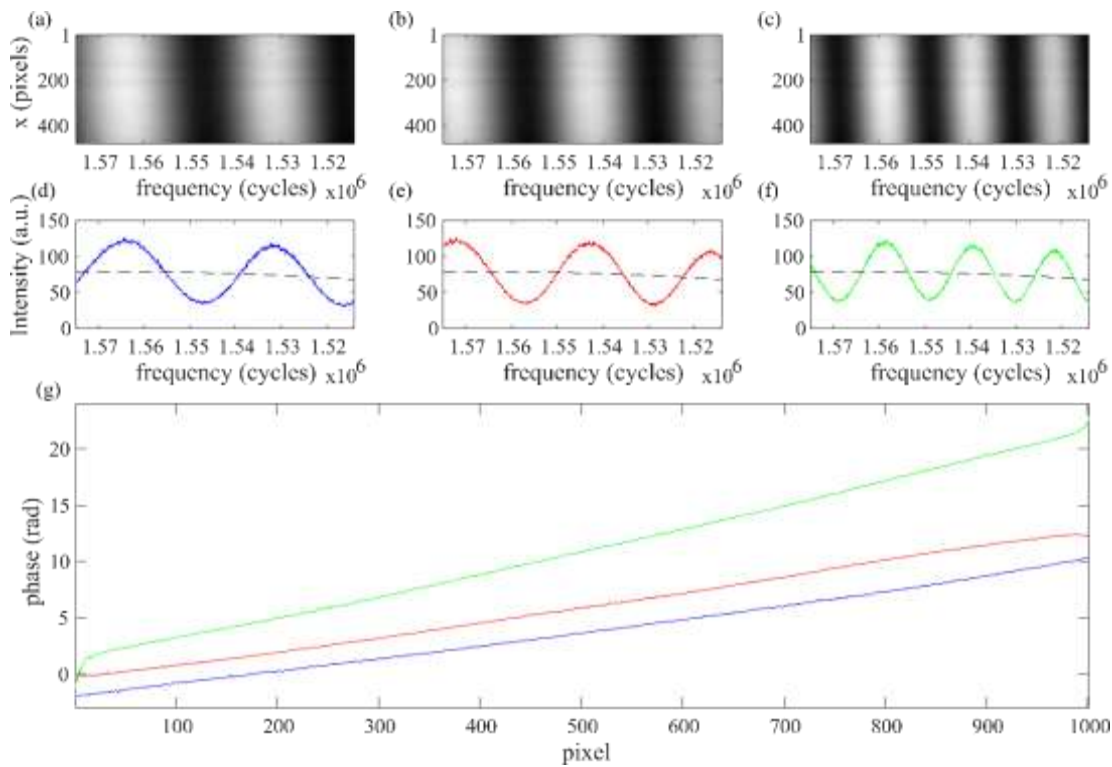

Figure 6: Parts (a), (b) and (c) show measurement results obtained by the instrument shown in fig (5) when a nominally flat object is measured at 3 locations close to equal optical path lengths in the reference and measurement arms. Parts (d), (e) and (f) show the intensity measured on the x = 100 row of pixels for the results shown in parts (a), (b) and (c) respectively. The dashed line in parts (a), (b) and (c) is the estimate of the a{x} term. Part (g) shows the unwrapped phase recovered phase once the a{x} estimate has been subtracted. It can be seen that, apart from at the ends where the highest errors are expected to be found, the unwrapped phase is close to a straight line, without the oscillations that would be found if a{x} was not subtracted.

Figure 6 parts (a-c) show the signal recorded when a nominally flat object is measured at three locations close to the point of equal optical path in the reference and measurement arms. Parts (d-f) show the signals that were recorded along the $x = 100$ row for the measurements shown in parts (a-c) respectively. The low number of oscillations present shows that the optical path length is nearly equal in the reference and measurement arms. The dashed black line in parts (d-e) is the same for all plots and corresponds to the estimation of $a\{x\}$. This estimate has been obtained by finding the locations of several maxima and minima on several signals and fitting a forth order polynomial to find an upper and lower envelope. $a\{x\}$ is then taken to be the average of these two envelopes.

The background signal, $a\{x\}$, is subtracted from the signal, before the phase is calculated in the manner described in the previous sections. The unwrapped phase for parts (d-f) is shown in part (g) with the colour of the line matching the colour of the line showing the data in parts (d-f). As can be seen, the unwrapped phase takes the form suggested in the previous section, unwrapping to almost a straight line with a slight ripple on it and more significant errors being found in the phase determination at the beginning and end of the array.

**Discussion**

In the preceding sections we have shown that the phase for the signal can be recovered with quite a high degree of accuracy as long as a few steps can be followed. We have shown that

- The background signal must be removed before the signal is Fourier transformed. This should be done so that the upper and lower bounds of the envelope are equal about the $y$ axis. Failure to do this will lead to a mixing of the phase associated with the signal we are interested in and the phase of the background signal, and can lead to significant errors. Alternatively, the Fourier transform of the background signal must be known well enough that it can be subtracted from the Fourier transform of the full signal, which achieves the same thing. Uncharacterised, or varying, background signals that have magnitudes that are not small compared to the interference fringes are likely to lead to significant errors.
- The envelope on the signal, $b\{x\}$, should be assessed by Fourier transforming it. If the elements of its Fourier transform have significant magnitude over a broad range of spatial frequencies an additional window should be applied to the data, to lead to a combined window with a few significant spatial frequency terms that occur at low spatial frequencies.
- The lowest frequency signal that the method should be used on corresponds to the location of the first significant element of the Fourier transform of the envelope. Below this frequency the phase of the envelope will significantly mix with the phase of the signal.

If this is carried out the method outlined in Takeda *et al*[1] should allow a reasonably accurate estimation of the phase even at low frequencies. Though in cases where high accuracy is needed, such as metrological applications, an assessment of the bounds on the error should be made for the specific window function on the signal. This can be carried out by following the example we illustrated for the case where $b\{x\} = 1$ within the measurement window.

Even following this route some errors will still occur, especially at very low frequencies, and this is due to the mixing of spatial frequencies on the positive and negative spatial frequency axis. It is possible that phase could be recovered through the use of multiple windows and comparing the different way these affect the results, however that is beyond the scope of what we wish to present here.

**Materials and methods**

The instrument used to generate the experimental results consists of a halogen lamp, the light from which is collected and coupled into a multimode fibre (600 $\mu m$ core diameter, NA 0.39). The light from each point at the end of the fibre is collimated, redirected using a non-polarising 50:50 beamsplitter before being incident on a Michelson interference objective (4x, WD: 27 mm). After the scattered light is collected by this objective it is imaged using a 200 mm tube lens (f = 250mm) onto a slit (300 $\mu m$) before being collimated and separated by wavelength by a reflective holographic diffraction grating with 1200 lines per mm. The light is then focussed by a lens (f = 60mm) onto a camera (IMPERX, Bobcat B0620, with 480 by 640 pixels, 207 fps) to create the signal.

## Acknowledgements

The authors gratefully acknowledge the Royal Academy and Renishaw PLC for their sponsorship of the Research Chair in Precision Metrology and the UK's Engineering and Physical Sciences Research Council (EPSRC) funding of the Future Metrology Hub (Grant Ref: EP/P006930/1).

## Conflict of interests

The authors declare they have no conflicts of interest.

## Contributions

The study was conceived by AH, DT and XJ. XJ oversaw the project, and the mathematical analysis was carried out by AH with input from XJ. DT developed the experimental setup and provided the experimental results.